\documentclass[a4paper,twoside,10pt]{article}

\input{jgrg20.sty}

\def\v0{{\bf 0}}

\begin{document}
%
\pagestyle{fancy}
\fancyhead{}
  \fancyhead[RO,LE]{\thepage}
  \fancyhead[LO]{N. Kan,~T. Maki, and K. Shiraishi}          
  \fancyhead[RE]{Weyl invariant DBI-Einstein theory}    
\rfoot{}
\cfoot{}
\lfoot{}
\label{P23}                             
\title{%
  Weyl invariant Dirac-Born-Infeld-Einstein theory
}
%
\author{%
  Nahomi Kan\footnote{Email address: kan@yamaguchi-jc.ac.jp}$^{(a)}$,
  Takuya Maki\footnote{Email address: maki@jwcpe.ac.jp}$^{(b)}$,
  and
  Kiyoshi Shiraishi\footnote{Email address: shiraish@yamaguchi-u.ac.jp}$^{(c)}$
}
%
\address{%
  $^{(a)}$Yamaguchi Junior College, Hofu-shi, Yamaguchi 747--1232, Japan\\
  $^{(b)}$Japan Women's College of Physical Education, 
Setagaya,~Tokyo 157-8565,~Japan\\ 
  $^{(c)}$Yamaguchi University, Yamaguchi-shi, Yamaguchi 753--8512, Japan\\}
%
\abstract{
We consider a Weyl invariant extension of Dirac-Born-Infeld type gravity.
An appropriate choice of the metric hides the scalar degree of freedom
which is required by the local scale invariance of the action at the first sight, 
and then a vector field acquires mass. 
Moreover, nonminimal couplings of the vector field and curvatures are induced, 
which may be suitable to the vector inflation scenario.
}

\section{Introduction}
The cosmological inflation is proposed as some resolutions for the important cosmological problems, 
{\it e.g.} the flatness, horizon and monopole problems. 
Most of successful models are based on models of classical scalar fields, 
although we have not known the reason of the existence of such fields
in the theory of elementary particle physics. 
Another inflation scenario, which is called the vector inflation,
is proposed by Ford~\cite{P23_F} and some authors~\cite{P23_BL,P23_GMV,P23_GV}. 
In such recent models~\cite{P23_GMV,P23_GV}, 
the massive vector field couples non-minimally to gravity.
The Lagrangian density  in the model~\cite{P23_GMV} is  expressed as 
\begin{equation}
{\cal L}=\sqrt{-g}\left[ \frac{R}{16\pi}-\frac{1}{4}F_{\mu\nu}F^{\mu\nu}
-\frac{1}{2}\left( m^2-\frac{R}{6} \right) A_\mu A^\mu \right]\,,
\label{P23_VI_Lagrangian}
\end{equation}
where $F_{\mu\nu}=\partial_\mu A_\nu - \partial_\nu A_\mu$
and the Newton constant is unity ($G=1$).
If we assume the spatially flat universe with the metric
\begin{equation}
ds^2=-dt^2+a^2(t) d\bf{x}^2\,,
\label{P23_VI_metric}
\end{equation}
and $A_i ~(i=1, 2, 3)$ depends only on $t$
and $A_0 =0$,
then the equation of motion of the vector field $A_i$ becomes the following form
\begin{equation}
\ddot{B}_i +3\frac{\dot{a}}{a}\dot{B}_i+m^2B_i=0\,,
\label{P23_VI_eqm_B}
\end{equation}
where $B_i \equiv A_i/a$.
Eq.~(\ref{P23_VI_eqm_B}) is very similar to the one 
for a homogeneous scalar field
in the Friedmann-Lema\^{i}tre-Robertson-Walker universe. Moreover, the energy
density is expressed as $\sim \dot{B}_i^2$, which is also similar to
the one for the scalar field. Thus the approximately isotropic ansatz can
be justified.

We have studied the vector inflation scenario~\cite{P23_MNS}
with Weyl invariance~\cite{P23_WG}.
We found that the choice of the frame yields the mass of the Weyl gauge field, 
but the nonminimal coupling term is lost~\cite{P23_MNS}.
We need to generalize further the gravitation theory. 

Dirac-Born-Infeld-Einstein (DBIE) theory was considered by Deser and Gibbons~\cite{P23_DG}
and have been studied by many authors~\cite{P23_DG2}.
The Lagrangian density of  DBIE theory
takes the following type 
\begin{equation}
{\cal L}\sim
\pm\sqrt{-\det(g_{\mu\nu}\pm\alpha R_{\mu\nu})}\,,
\label{P23_origDG}
\end{equation}
where $R_{\mu\nu}$ is the Ricci tensor and the $\alpha$ is a constant.
Originally, electromagnetism of the Dirac-Born-Infeld (DBI) type has been
considered as a candidate of the nonsingular theory of electric fields.
Therefore the DBIE theory as the highly-nonlinear theory is also expected
as a theory of gravity suffered from no argument of singularity.

We take notice of the nonlinearity in the DBIE theory and
expect that the Weyl invariant DBIE theory realizes the suitable scenario of the cosmological inflation.

\section{Weyl's gauge gravity theory}
In this section, we briefly review the Weyl's gauge transformation
to construct the gauge invariant Lagrangian.
Consider the transformation of metric in $D$ dimensions  
\begin{equation}
g_{\mu\nu}\rightarrow g'_{\mu\nu}=e^{2\Lambda(x)}g_{\mu\nu}\,,
\label{P23_trans_metric1}
\end{equation}
where $\Lambda(x)$ is an arbitrary function of the coordinates $x^\mu$.
We can define the field with weight $d=-\frac{D-2}{2}$ which transforms as
\begin{equation}
\Phi\rightarrow \Phi'=e^{-\frac{D-2}{2}\Lambda(x)}\Phi\,.
\end{equation}
In order to construct the locally invariant theory, 
we consider the covariant derivative of the scalar field  
\begin{equation}
\tilde\partial_\mu\Phi\equiv
\partial_\mu\Phi-\frac{D-2}{2}\,A_\mu\Phi\,,
\end{equation}
where $A_\mu$ is a Weyl's gauge invariant vector field.
Under the Weyl gauge field transformation
\begin{equation}
A_\mu\rightarrow A'_\mu=
A_\mu-\partial_\mu\Lambda(x)\,,
\label{P23_gaugetrans}
\end{equation}
we obtain the transformation of the covariant derivative of the 
scalar field as
\begin{equation}
\tilde\partial_\mu\Phi\rightarrow
e^{-\frac{D-2}{2}\Lambda(x)}\tilde\partial_\mu\Phi\,.
\end{equation}
The modified Christoffel symbol and the modified curvature
are given as follows 
\begin{equation}
\tilde\Gamma^\lambda_{\mu\nu}\equiv\frac{1}{2}g^{\lambda\sigma}
\left(\tilde\partial_\mu
g_{\sigma\nu}+\tilde\partial_\nu
g_{\mu\sigma}-\tilde\partial_\sigma
g_{\mu\nu}\right)\,,
\end{equation}
\begin{equation}
\tilde{R}^\mu{}_{\nu\rho\sigma}[g, A]\equiv
\partial_\rho\tilde\Gamma^\mu_{\nu\sigma}-
\partial_\sigma\tilde\Gamma^\mu_{\nu\rho}+
\tilde\Gamma^\mu_{\lambda\rho}\tilde\Gamma^\lambda_{\nu\sigma}-
\tilde\Gamma^\mu_{\lambda\sigma}\tilde\Gamma^\lambda_{\nu\rho}\,,
\end{equation}
where
$\tilde{\partial}_\mu g_{\sigma\mu}\equiv{\partial}_\mu g_{\sigma\mu}+2A_\mu g_{\sigma\mu}$.
The Ricci curvature is generalized as 
\begin{eqnarray}
\tilde{R}_{\nu\sigma}[g,A]
&\equiv&
\tilde{R}^\mu{}_{\nu\mu\sigma}[g,A]\nonumber \\
&=&R_{\nu\sigma}+F_{\nu\sigma}-
\left[(D-2)\nabla_\sigma A_\nu+g_{\nu\sigma}\nabla_\mu A^\mu \right]
+(D-2)\left(A_\nu A_\sigma-A_\lambda A^\lambda g_{\nu\sigma}\right)\,,
\label{P23_WRicci}
\end{eqnarray}
where 
the field strength of the vector field is given by
$
F_{\mu\nu}\equiv\partial_\mu A_\nu-\partial_\nu A_\mu\,,
$
which is gauge invariant as
$
F_{\mu\nu}\rightarrow F'_{\mu\nu}=F_{\mu\nu}\,.
$
Note that under the Weyl's gauge transformation,
The Ricci curvature (\ref{P23_WRicci})
is invariant 
\begin{equation}
\tilde{R}_{\nu\sigma}[g,A]\rightarrow\tilde{R}_{\nu\sigma}[g',A']=
\tilde{R}_{\nu\sigma}[g,A]\,.
\end{equation}

\section{Weyl invariant Dirac-Born-Infeld gravity}
We can use the Weyl invariant Ricci tensor $\tilde{R}_{\mu\nu}$ in the DBI gravity.
We should also use a combination 
$\Phi^\frac{4}{D-2}g_{\mu\nu}$
instead of the metric tensor, because it is not Weyl invariant. 
Note that the scalar $\Phi$ compensates the dimensionality of the metric. 
The use of $\tilde{R}_{\mu\nu}$ and  $\Phi^\frac{4}{D-2}g_{\mu\nu}$ 
in the DBI type action leads to the theory of gravity, a vector field,
and unexpectedly, a scalar field.

Now
we introduce the following independently Weyl invariant tensors
into the determinant in the DBI theory
\begin{eqnarray}
& &\Phi^\frac{4}{D-2}g_{\nu\sigma}\,,\quad
\tilde{R}^S_{\nu\sigma}[g,A]\,,\quad
\tilde{R}[g,A]g_{\nu\sigma}\,,\quad
F_{\nu\sigma}\,,\quad
\Phi^{-2}\tilde{\partial}_\nu\Phi\tilde{\partial}_\sigma\Phi\,,\nonumber
\\ & &\Phi^{-2}g^{\lambda\mu}\tilde{\partial}_\lambda\Phi
\tilde{\partial}_\mu\Phi g_{\nu\sigma}\,,\quad
\tilde{\nabla}_\sigma(\Phi^{-1}\tilde{\partial}_\nu\Phi)
+\tilde{\nabla}_\nu(\Phi^{-1}\tilde{\partial}_\sigma\Phi)
\,,\quad
\tilde{\nabla}^\mu(\Phi^{-1}\tilde{\partial}_\mu\Phi)
g_{\nu\sigma}\,,
\label{P23_Wtensors}
\end{eqnarray}
where
\begin{equation}
\tilde{R}^S_{\nu\sigma}[g,A]
=R_{\nu\sigma}-\left[
\frac{D-2}{2}(\nabla_\sigma A_\nu+\nabla_\nu A_\sigma)+
g_{\nu\sigma}\nabla_\mu A^\mu
\right]+(D-2)\left(
A_\nu A_\sigma
-A_\lambda A^\lambda
g_{\nu\sigma}\right)\,,
\end{equation}
and 
\begin{equation}
\tilde{R}[g,A]\equiv
g^{\nu\sigma}\tilde{R}_{\nu\sigma}[g,A]=R-2(D-1)\nabla_\mu
A^\mu-(D-1)(D-2)A_\mu A^\mu\,.
\label{P23_R}
\end{equation}
We choose those as symmetric tensors are not traceless.%
\footnote{Judging from the number of fields and derivatives,
the term $\Phi^{-\frac{4}{D-2}}g^{\lambda\mu}F_{\nu\lambda}F_{\sigma\mu}$
is allowed in the same order. But this term is different from others in
the point that it includes two kinds of fields except for the metric. We
discarded this marginally possible term here.}
Using the Weyl invariant tensors (\ref{P23_Wtensors}),
 our model 
 is described by the Lagrangian density
\begin{equation}
{\cal L}=-\sqrt{-\det M_{\mu\nu}}+(1-\lambda)\sqrt{-\det
(\Phi^{\frac{4}{D-2}}g_{\mu\nu})}\,,
\label{P23_WDBIE1}
\end{equation}
with
\begin{eqnarray}
M_{\mu\nu}&\equiv&\Phi^{\frac{4}{D-2}}g_{\mu\nu}-\alpha_1
\tilde{R}^S_{\mu\nu}[g,A]-\alpha_2
\tilde{R}[g,A]g_{\mu\nu}+\beta F_{\mu\nu}\nonumber \\& &
+
\gamma_1\Phi^{-2}\tilde{\partial}_\mu\Phi\tilde{\partial}_\nu\Phi
+\gamma_2\Phi^{-2}g^{\lambda\sigma}\tilde{\partial}_\lambda\Phi
\tilde{\partial}_\sigma\Phi g_{\mu\nu}\nonumber \\
& &-\gamma_3\left[\tilde{\nabla}_\mu(\Phi^{-1}\tilde{\partial}_\nu\Phi)
+\tilde{\nabla}_\nu(\Phi^{-1}\tilde{\partial}_\mu\Phi)\right]
-\gamma_4\, g^{\lambda\sigma}
\tilde{\nabla}_\lambda(\Phi^{-1}\tilde{\partial}_\sigma\Phi)
g_{\mu\nu}\,,
\label{P23_M1}
\end{eqnarray}
where
$\alpha_1$, $\alpha_2$, $\beta$, $\gamma_1$, $\gamma_2$,
$\gamma_3$, $\gamma_4$ and $\lambda$ are dimensionless constants.%
\footnote{
If we demand that the terms with lowest derivatives 
in the expansion of the Weyl invariant Lagrangian density (\ref{P23_WDBIE1}) 
look like the one 
of scalar-tensor theory,
we must choose as
$\alpha_1+4\alpha_2>0$ and $\gamma_1+4\gamma_2+4\gamma_3+8\gamma_4>0$, for
$D=4$.}
Furthermore the Lagrangian density can be expressed by the
new metric conformally related to the original one and new variables.
If 
we choose
\begin{equation}
\hat{g}_{\mu\nu}\equiv f^{-2}\Phi^{\frac{4}{D-2}}g_{\mu\nu}\,,
\end{equation}
where a mass scale $f$ was introduced, 
and 
\begin{equation}
\hat{A}_\mu\equiv A_\mu-\frac{2}{D-2}\partial_\mu\ln\Phi\,,
\end{equation}
we can rewrite the Lagrangian density (\ref{P23_WDBIE1}) as
\begin{equation}
{\cal L}=-\sqrt{-\det \hat{M}_{\mu\nu}}+(1-\lambda)f^D\sqrt{-\hat{g}}\,,
\label{P23_WDBIE2}
\end{equation}
where $\hat{g}=\det \hat{g}_{\mu\nu}$ and
{
\begin{equation}
\hat{M}_{\mu\nu}=
f^{2}\hat{g}_{\mu\nu}-\alpha_1{\tilde{R}}_{\mu\nu}
-\alpha_2{\tilde{R}}\hat{g}_{\mu\nu}+\beta \hat{F}_{\mu\nu}
+\gamma'_1\hat{A}_\mu \hat{A}_\nu
+\gamma'_2 \hat{g}^{\rho\sigma}\hat{A}_\rho
\hat{A}_\sigma \hat{g}_{\mu\nu}
+\gamma'_3\left({\hat{\nabla}}_\mu \hat{A}_\nu
+{\hat{\nabla}}_\nu \hat{A}_\mu\right)
+\gamma'_4\, 
{\hat{\nabla}}^\rho \hat{A}_\rho\,
\hat{g}_{\mu\nu}\,,
\label{P23_M2}
\end{equation}
}
in which $\gamma'_1$, $\gamma'_2$, $\gamma'_3$ and $\gamma'_4$ are dimensionless constants 
rewritten by the set of the former, 
$\alpha_1$, $\alpha_2$, $\gamma_1$, $\gamma_2$,
$\gamma_3$ and $\gamma_4$.
Note that the scalar field $\Phi(x)$ is hidden away in the Lagrangian (\ref{P23_WDBIE2}).
The expansion of the determinant in (\ref{P23_WDBIE2})
yields the non-minimal coupling term to gravity of the vector field
and the induced curvature term
as well as the ordinary scalar curvature and 
the gauge sector.%
\footnote{The expansion of the determinant is known for various dimensions.
}  
Therefore 
(\ref{P23_WDBIE2})
is the promising Lagrangian for describing the vector inflation.

\section{Cosmology of Weyl's gauge gravity}
We consider cosmological aspect of the theory described by the
Lagrangian (\ref{P23_WDBIE2}). 
We assume the four dimensional flat universe
and take the isotropic metric 
\begin{equation}
ds^2=-dt^2+a^2(t)d\bf{x}^2\,. 
\end{equation}
We also assume 
that only $A_1(t)$ is homogeneously evolving, and $A_2=A_3=A_0=0$.
By these ansatze, we look for the condition that the vector field 
behaves much like a scalar field at classical homogeneous level.
Substituting the ansatze into  (\ref{P23_M2}),
and after some calculation,
we can extract the part of the Lagrangian which
includes bilinear and higher-order of the vector field $A_1$.
If we choose the parameters as
$\beta^2=\frac{1}{2}\left(5\alpha_1\gamma'_1+12\alpha_2\gamma'_1+
12\alpha_1\gamma'_2+48\alpha_2\gamma'_2\right)$ 
and
$(\gamma'_3)^2=-\frac{1}{2}\alpha_1\gamma'_1\,,$
we find that 
the vector-field part becomes 
\begin{equation}
a^3\left[\frac{1}{2}(\beta^2-{\gamma_3'}^2)\dot{B}_1^2-
\frac{f^2}{2}(\gamma_1'+4\gamma_2')B_1^2-
\frac{1}{8}\left(-{\gamma_1'}^2+4\gamma_1'\gamma_2'+8{\gamma_2'}^2\right)B_1^4+\cdots
\right]\,,
\end{equation}
where $B_1=\frac{A_1}{a}$,
which acts as a scalar field.

We can tune the parameters
to obtain the model of the realistic inflation.
If $\alpha_2=\gamma_1=\gamma_2=\gamma_3=\gamma_4=0$, 
or these parameters take small values in comparison with $\alpha_1$,
there is only one parameter $\alpha_1$ in the model. 
Unfortunately,
the effective mass for $B_1$ may be large in this simple model.
Another tuning of the parameters is also possible. 
If we choose appropriate parameters,
the chaotic inflation~\cite{P23_chaotic} 
is practicable  
via the self-interaction of the scalar field $B_1$.

\section{Summary and Outlook}
We investigated Weyl invariant  Dirac-Born-Infeld gravity.
The choice of an appropriate frame breaks the Weyl invariance, 
and the vector field acquires mass
as well as non-minimal coupling to gravity,
and curvatures are induced. 
Therefore 
the Weyl invariant DBIE theory 
is expected to be
a candidate for a model which causes an inflationary universe. 
We also examined 
slow development of the massive vector field
and indicated that
several scenarios of the inflation
are possible
by  tuning of parameters appropriately.

Future works are in order.
Numerical calculation and large simulation will be needed 
to understand the minute meaning of the Weyl invariant DBI gravity, 
because of the local inhomogenuity of the direction 
as well as the strength of vector fields is important for thorough understanding.%
\footnote{The calculation will be slightly simplified if we use a scalar auxiliary field.}      
The inflation along with a fast evolution, known as the DBI inflation,
is also interesting.
The similar scenario is feasible in our model, 
though the higher-derivatives make the detailed analysis difficult.
The higher-dimensional cosmology in the Weyl invariant DBI gravity is worth studying 
because of its rich content.
Incidentally, DBI gravity in three dimensions is eagerly studied, 
which is related to new massive gravity theory. 
We think that the Weyl invariant extension is also of much interest.

\section*{Acknowledgements}
The authors would like to thank K. Kobayashi for useful comments, and also the organizers of JGRG20.


\end{document}